\documentclass[12pt]{article}
\usepackage{graphicx}
\usepackage{color}
\usepackage{amsmath}
\usepackage{epsfig}
\usepackage{comment}
\usepackage{latexsym}
\usepackage{color}
\newcommand{\mysquare}[0]{\raise-.2ex\hbox{{\Large$\Box$}}}
\def\lsim{\mathrel{\rlap {\raise.5ex\hbox{$ < $}}
{\lower.5ex\hbox{$\sim$}}}}
\def\gsim{\mathrel{\rlap {\raise.5ex\hbox{$ > $}}
{\lower.5ex\hbox{$\sim$}}}} \topmargin -1.5cm \textheight=22.5cm
\catcode`\@=11
\newcount\hour
\newcount\minute
\newtoks\amorpm
\hour=\time\divide\hour by60 \minute=\time{\multiply\hour by60
\global\advance\minute by-\hour}
\edef\standardtime{{\ifnum\hour<12 \global\amorpm={am}%
        \else\global\amorpm={pm}\advance\hour by-12 \fi
        \ifnum\hour=0 \hour=12 \fi
        \number\hour:\ifnum\minute<10 0\fi\number\minute\the\amorpm}}
\edef\militarytime{\number\hour:\ifnum\minute<10 0\fi\number\minute}
\def\draftlabel#1{{\@bsphack\if@filesw {\let\thepage\relax
   \xdef\@gtempa{\write\@auxout{\string
      \newlabel{#1}{{\@currentlabel}{\thepage}}}}}\@gtempa
   \if@nobreak \ifvmode\nobreak\fi\fi\fi\@esphack}
        \gdef\@eqnlabel{#1}}
\newcommand{\be}[0]{\begin{equation}}
\newcommand{\ee}[0]{\end{equation}}
\newcommand{\ba}[0]{\begin{eqnarray}}
\newcommand{\ea}[0]{\end{eqnarray}}

\def\ea{\end{eqnarray}}
\def\bs{\begin{subequations}}
\def\es{\end{subequations}}

\def\thefootnote{\fnsymbol{footnote}}
\def\es{\end{subequations}}

\def\ec{\hat E^{c}_{2}}

\def\nn{\nonumber}

\def\be{\begin{equation}}
\def\ee{\end{equation}}
\def\bc{\begin{center}}
\def\ec{\end{center}}
\def\bea{\begin{eqnarray}}
\def\eea{\end{eqnarray}}

\def\nn{\nonumber}
\usepackage{amsmath}
\usepackage{amssymb,amsfonts}
\usepackage{stmaryrd}
\newcommand{\pd}{\partial}

\begin{document}
%\verb|\usepackage{draftcopy}|\\
%\renewcommand{\theequation}{\arabic{section}.\arabic{equation}}
%
%\renewcommand{\section}{\setcounter{equation}{0}\@startsection%
%{section}{1}{0mm}{-\baselineskip}{0.5\baselineskip}%
%{\normalfont\normalsize\bfseries}}
%
%\renewcommand{\subsection}{\@startsection%
%{subsection}{2}{0mm}{-\baselineskip}{0.5\baselineskip}%
%{\normalfont\normalsize\slshape}}
\begin{titlepage}
\begin{flushright}
LPTENS--07/16\\
%hep-th/07xxxxx \\
April 2007
\end{flushright}

\vspace{8mm}

\begin{centering}
{\bf\LARGE A Wave-function for Stringy Universes$^\ast$\\

\vspace{9mm}
 {\large Costas~Kounnas,$^{1}$ Nicolaos
 Toumbas$^2$ and Jan Troost$^1$}

\vspace{6mm}

{\small $^1$ Laboratoire de Physique Th\'eorique,
Ecole Normale Sup\'erieure,$^\dagger$ \\
24 rue Lhomond, F--75231 Paris Cedex 05, France

\vskip .1cm

$^2$ Department of Physics, University of Cyprus, Nicosia 1678,
Cyprus.
\\ }

\vspace{9mm}

{\bf\Large Abstract}}

\end{centering}

\vspace{4mm}

\begin{quote}
We define a wave-function for string theory cosmological
backgrounds. We give a prescription for computing its norm
following an earlier analysis within general relativity. Under Euclidean
continuation, the cosmologies we discuss in this paper are described
in terms of compact parafermionic worldsheet systems. To define the
wave-function we provide a T-fold description of the parafermionic
conformal field theory, and of the corresponding string cosmology.
In specific examples, we compute the norm of the wave-function and
comment on its behavior as a function of moduli.

\end{quote}

\vspace{5pt} \vfill
% \begin{flushleft}
% LPTENS--07/xx\\
% April 2007
% \end{flushleft}
\hrule width 6.7cm \vskip.1mm{\small \small \small \noindent
$^\ast$\ Research partially supported by the EU under the contracts
MRTN-CT-2004-005104, MRTN-CT-2004-512194 and ANR (CNRS-USAR)
contract No 05-BLAN-0079-01 (01/12/05).\\
$^\dagger$\ Unit{\'e} mixte  du CNRS et de l'Ecole Normale
Sup{\'e}rieure associ\'ee \`a l'universit\'e Pierre et Marie Curie 6,
UMR 8549.}
\end{titlepage}
\newpage
\setcounter{footnote}{0}
\renewcommand{\thefootnote}{\arabic{footnote}}

\setlength{\baselineskip}{.7cm} \setlength{\parskip}{.2cm}

%% Section 1:
%\setcounter{section}{0}
\section{Introduction}
Our goal in this paper is to embed the Hartle-Hawking no-boundary
proposal for a wave-function description of the quantum state of the
universe \cite{Hartle:1983ai}\cite{Vilenkin:1984wp}
 in a string theoretic framework. The
Hartle-Hawking proposal pertains in particular to de Sitter--like
universes in general relativity. A partial list of interesting
recent work on related topics is
\cite{Ooguri:2005vr}\cite{Sarangi:2005cs}\cite{Brustein:2005yn}
\cite{Cardoso:2006nt}\cite{Sarangi:2006eb}\cite{Barvinsky:2006uh}.

One motivation for embedding the Hartle-Hawking proposal
into string theory is that it provides us with a calculable
quantity in de Sitter-like compactifications of a quantum theory of gravity.
 These quantities are hard
to come by (see e.g. \cite{Antoniadis:1985pj}\cite{Banks:2005bm} and
references
thereto).

Two essential properties which a cosmological background must
fulfill in order to admit a wave-function description under the
no-boundary proposal are the following. First the cosmology must be
spatially closed. More importantly, the cosmology should admit a
continuation to a positive definite Euclidean geometry that is
compact and has no boundaries or singularities. The most
familiar example is the case of $n$-dimensional de Sitter space,
$dS_n$, where these properties are satisfied.
In global coordinates, the $dS_n$ metric is given by
\begin{equation}
ds^2=R^2(-dt^2+\cosh^2 t ~d\Omega^2),
\end{equation}
where $d\Omega^2$ is the metric on a round unit $(n-1)$-sphere and
$R$ is the radius of curvature. The spatial slices of constant time $t$
are $(n-1)$-spheres of radius $R\cosh t$. We can rotate to Euclidean
signature by setting $t = i\tau=i (\pi/2 -\theta) $,
upon which we obtain an $n$-sphere $S_n$ of radius $R$. The
Euclidean continuation is a compact smooth manifold.

In a field theoretic framework, the quantum state of a de Sitter
cosmology can be expressed as a functional of fields, including both
matter fields and metric fluctuations, on a spatial slice of
time-reversal symmetry. For the reversal $t \rightarrow -t$,
this is the slice $t=0$ in the de
Sitter space $dS_n$. That slice is also the equator $\theta=\pi/2$
of the corresponding Euclidean sphere $S_n$. Imagine cutting de
Sitter space along this slice and gluing smoothly one half of it to
half a sphere $S_n$. Under the Hartle-Hawking proposal, we
express the wave-function as a Euclidean path integral over half the
sphere $S_n$ with the condition that the metric $g_{ij}$ and the
matter fields, collectively denoted by $\phi$, take specific values
$(h_{ij}, \phi_0)$ on the boundary equator $\theta=\pi/2$:
\begin{equation}
\Psi(h_{ij}, \phi_0)=\int [dg][d\phi] e^{-S_E(g, \phi)}.
\end{equation}
No other boundary condition needs to be specified due to the
compactness of the Euclidean manifold. Here, $S_E$ is the Euclidean
gravitational action in the presence of matter fields and a positive
cosmological constant $\Lambda$.

The norm of the wave-function is given by the full Euclidean path
integral on $S_n$. It can be computed in the semi-classical
approximation by evaluating the Euclidean action for a given
solution to the classical equations of motion. One solution is empty
de Sitter space of radius $R \sim \Lambda^{-1/2}$. In this
approximation, and in the case of four dimensions, the norm is given
by  \cite{Hartle:1983ai}:
\begin{equation}
||\Psi_{HH}||^2\sim e^{2 \over 3\lambda}~,
\end{equation}
where the dimensionless parameter $\lambda$ is proportional to the
cosmological constant:
$$\lambda=2G\Lambda/9\pi ~.$$

The compactness of the Euclidean manifold ensures that the full path
integral is free of any infrared divergences. However the field
theory in question is non-renormalizable, and to go beyond the
semi-classical approximation, we need to impose an ultraviolet
cutoff. One way to deal with the ultraviolet ambiguities is to embed
the calculation in a string theoretic framework, where we expect the
ultraviolet divergences to be absent. Unfortunately there are no
known classical de Sitter solutions in string theory to begin with.
Therefore, we seek other cosmological backgrounds which are
exact solutions to string theory and for which we can generalize the
Hartle--Hawking computation.

To this end, notice that any tachyon free, compact Euclidean string
background provides us with a finite, calculable quantity,
namely the string partition function $Z_{string}$. Associated to the
classical string background is a two-dimensional worldsheet
conformal field theory (CFT). At the perturbative level, the string
partition function can be computed as usual as a sum of CFT vacuum
amplitudes over compact worldsheets of all topologies. Our proposal is
that when such Euclidean string backgrounds admit a continuation to
a Lorentzian cosmology, the Hartle-Hawking construction can be
generalized with the norm of the wave-function given by
\begin{equation}
||\Psi_{cosm.}||^2=e^{Z_{string}}.
\end{equation}
We will motivate this formula by working out specific examples in
string perturbation theory. As we will explain, the relevant string
partition function has to be thermal.

Given the discussion above, a first candidate to consider is a
Euclidean model for which the two-dimensional CFT is of the form $SU(2)_k\times K$, 
the first factor
corresponding to an $SU(2)$ Wess-Zumino-Witten (WZW) model at level
$k$ and the second factor $K$ corresponding to a suitable internal
compact conformal field theory. The WZW factor is equivalent to a sigma model on a
3-sphere of radius $(k \alpha')^{1/2}$ and with $k$ units of NSNS
3-form flux through the sphere. The dilaton field is constant and by
choosing this to be small we can apply string perturbation theory.
Unfortunately, however, the continuation to Lorentzian signature
results in a $dS_3$ cosmology with imaginary flux, and it is not
clear whether such a Lorentzian background is physical. (See 
\cite{Antoniadis:1988vi} for an alternative non-compact, time-like Liouville
model for which the $SU(2)$ WZW factor describes the internal space.)

The only known string theory example which satisfies all the
criteria we described so far is based on the parafermionic
$SU(2)_{|k|}/U(1)$ coset model \cite{Fateev:1985mm},
 which can be realized as a gauged
$SU(2)$ WZW model at level $|k|$. We consider Euclidean backgrounds
corresponding to a two-dimensional CFT of the form
$SU(2)_{|k|}/U(1)~\times ~K $ where $K$ is again an internal
compact conformal field theory. Such a Euclidean background admits a
Lorentzian continuation to a cosmological background belonging to a
class of models studied in
\cite{kounnas-lust}\cite{cornalba}, and which are described by
two-dimensional CFTs of the form $SL(2, R)_{-|k|}/U(1)~\times ~K$.
To avoid having to deal with the tachyonic instabilities of bosonic
string theory, we consider solutions of this form in superstring
theory.  The total central charge must
be $c_{tot}=15$ ($\hat c=10$) in order for worldsheet 
(super-)gravitational anomalies to cancel. When we fix the internal conformal field
theory
$K$, the level $|k|$ is determined by anomaly
cancellation.

The non-trivial time-dependence of the cosmology necessarily
breaks space-time supersymmetry. As in the de Sitter case, the
Euclidean path integral can be interpreted as a thermal ensemble.
Thus from the point of view of the
Euclidean $N=2$ worldsheet superconformal system,
 space-time
supersymmetry will
be broken by specific boundary conditions, analogous to the thermal co-cycles
that appear in the partition function
of superstring theories on flat space at finite temperature \cite{Atick:1988si}. For
large level $|k|$, the effective temperature of the models is of order $T
\sim 1/ \sqrt{|k|\alpha'}$ \cite{cornalba}\cite{Toumbas:2004fe}. In this paper, we will
explore some low level $|k|$ models. 
In order for the corresponding cosmological
wave-function to be computable in string perturbation theory,
the effective temperature must be below the Hagedorn temperature. 
A Hagedorn temperature would signal
a phase transition, as proposed in
\cite{Atick:1988si}\cite{Antoniadis:1991kh}\cite{Antoniadis:1999gz}.
We will construct explicitly low
level $|k|$ models for which the effective temperature is below the
Hagedorn temperature and so string perturbation theory can be applied.

It is well known that the geometric sigma model approach to the
parafermionic coset model (and to the corresponding Lorentzian
cosmology) leads to a metric with curvature singularities and strong
coupling. However, the underlying CFT is perfectly well behaved at
these apparently singular regions, and by using T-duality a weakly
coupled description of these regions can be
obtained \cite{kiritsis-kounnas}. Using this fact, we construct an
almost geometrical description of the CFT in terms of a compact,
non-singular T-fold \cite{Hellerman:2002ax}\cite{Dabholkar:2002sy}
with a well--defined partition function. These
considerations allow us to define the wave-function of the Lorentzian
cosmology.

Our paper is organized as follows. In section \ref{cosmo}, we
review properties of the two-dimensional $SL(2,
R)_{-|k|}/U(1)~\times ~K$ conformal field theory that corresponds to a cosmological
background. It is the analogue of the de Sitter universe. In section
\ref{euclidean}, we describe how to analytically continue the
cosmology to a compact Euclidean space-time described at the string
level by a two-dimensional parafermionic model of the form
$SU(2)_{|k|}/U(1)~\times ~K $. Then, we discuss in section
\ref{tfold} how to obtain an almost geometrical description of these
backgrounds in terms of T-folds. We discuss in sections
\ref{wavefunction} and \ref{norm} how to calculate a wave-function
and its norm for the cosmology. In section 7 we discuss the thermal nature
of the wave-function. In section
\ref{models} we apply the definition of the wave-function to some particular 
compact models and for which perturbation theory can be used
to compute its norm.
Finally we
discuss interpretations of the results in the concluding sections.

\section{The cosmological solution}
\label{cosmo} In this section, we review in some detail the
cosmological solution of string theory which is based on an
$SL(2,R)/U(1)$ gauged WZW model at level $k$ \cite{kounnas-lust}.

We can define a WZW conformal field theory on the group manifold
$SL(2,R)$, at least classically. The worldsheet action is given by
\begin{equation}
S= {k \over 4\pi}\int_{\Sigma}d^2zTr(g^{-1}\partial g g^{-1}
\bar{\partial}g) + {ik \over 12} \int_M Tr(g^{-1}dg \wedge g^{-1}dg
\wedge g^{-1}dg) \label{action},
\end{equation}
where $\Sigma$ is the worldsheet Riemann surface, $M$ is a
3-manifold whose boundary is $\Sigma$ and $g$ is an element of
$SL(2,R)$.

For concreteness, we parameterize the $SL(2,R)$ group manifold as
follows
\be
g =
\begin{pmatrix}
 & a & u    \\
                & -v  & b
\end{pmatrix}
\label{parameter}
\ee with $ab+uv=1$. The conformal field theory has an $SL(2,R) \times SL(2,R)$ global
symmetry. We choose to gauge an axial $U(1)$ subgroup under which
$g\to hgh$. In particular, we consider the non-compact $U(1)$
subgroup generated by \be \delta g =\epsilon
\begin{pmatrix}
&1 & 0 \\
           &0 & -1
\end{pmatrix}
g+\epsilon g
\begin{pmatrix}
&1 & 0 \\
           &0 & -1
\end{pmatrix}.
\ee Infinitesimally, we have the transformations
 $\delta a =2\epsilon a$, $\delta b = -2\epsilon b$,
$\delta u=\delta v=0$. To gauge this $U(1)$ symmetry, we introduce
an Abelian gauge field
 and
render the action invariant. The action is quadratic and
non-derivative in the gauge field, and so this can be integrated out
in a straightforward way \cite{witten}.

In the region $1-uv>0$, we can use the gauge freedom to set $a=b$
and integrate out the gauge field. The resulting action is expressed
in terms of gauge invariant degrees of freedom only, and it turns out to
be \be S=-{k \over 4\pi}\int d^2z {\partial u \bar{\partial} v
+\partial v \bar{\partial}u \over 1-uv}, \ee while a non-trivial
coupling to the worldsheet curvature is generated corresponding to a
dilaton background \cite{witten}. This action can be identified with
a non-linear sigma-model action with background metric
\be ds^2=-k
\alpha' {du dv \over 1-uv}. \label{metric} \ee The non-trivial
dilaton is given by \be e^{2\Phi}={e^{2\Phi_0} \over 1-uv}.
\ee

The metric (\ref{metric}) is a Lorentzian metric whose precise
causal structure, however, depends on the sign of $k$. For positive
level $k$, $u$ and $v$ are Kruskal-like null coordinates of a
2-dimensional black hole. In this case, the time-like
coordinate is given by $u+v$, and the metric has space-like
singularities in future and past times at $uv=1$.

For negative level $k$, one obtains a cosmological solution
\cite{kounnas-lust}. It consists of a singularity-free light-cone
region, and there are (apparent) time-like singularities in the
regions outside the light-cone horizons. Indeed, for negative level
$k$ we may set $u=-T+X$ and $v=T+X$
%[$T=(v-u)/2$ is the time-like
%coordinate]
and the metric becomes \be ds^2=|k| \alpha'
{-dT^2+{dX}^2 \over 1+{T}^2-{X}^2}. \ee The surfaces of constant
time  $T$
intersect the singularities at $X = \pm \sqrt{1+T^2}$. Even though
the singularities follow accelerated trajectories, their proper
distance remains finite with respect to the string frame metric \be
L= (|k|\alpha')^{1\over 2}\int_{-\sqrt{1+T^2}}^{\sqrt{1+T^2}}{dX
\over \sqrt{1+T^2-X^2}}=\pi(|k|\alpha')^{1\over 2}. \ee
So with respect to stringy probes, the cosmology is spatially closed.

The singularity-free light-cone region is the region $T^2-X^2 \ge 0$
(or $uv \le 0$). The future part of this region describes an
expanding, asymptotically flat geometry with the string coupling
vanishing at late times. See e.g.
\cite{Antoniadis:1988vi}\cite{kounnas-lust}\cite{cornalba}\cite{Elitzur:2002rt}\cite{Giveon:2003gb}\cite{Berkooz:2002je}\cite{Strominger:2003fn}\cite{Hikida:2004mp}\cite{Toumbas:2004fe}\cite{Nakayama:2006gt}
for some discussions of these types of
models.
To see this, we parameterize the region $uv \le 0$ with
coordinates $(x,t)$ such that \be u=-t e^x, \,\,\, v=te^{-x} \ee and
the metric becomes \be ds^2=|k| \alpha' {-dt^2+ t^2{dx}^2 \over
1+t^2}, \ee while the dilaton field becomes \be
e^{2\Phi}={e^{2\Phi_0} \over 1+t^2}. \ee The scalar curvature is
given by
\begin{equation}
{\cal R} \sim {1 \over |k|\alpha'( 1+ t^2)}.
\end{equation}
Initially the curvature is set by the level $|k|$ and it is
positive. No matter how small the level $|k|$ is, asymptotically
%for large $t$, 
the scalar curvature vanishes.
An observer in this region
never encounters the singularities. These are hidden behind
the visible horizons at $T=\pm X$. However signals from the
singularities can propagate into the region $uv<0$, and therefore
influence its future evolution.

Thus when $|k|$ is small the early universe region $t \sim 0$ is highly
curved, with curvature of order the string scale. In this sense,
it is similar to a big-bang cosmology.
Despite the regions of large curvature,
this
cosmological
background has a well defined CFT description and can be described
in a string theoretic framework.

The cosmological background can also be realized as a solution of
superstring theory by generalizing the worldsheet theory to a
superconformal $SL(2,R)/U(1)$ model. The central charge of the
superconformal $SL(2,R)/U(1)$ model at negative level $k$, is given
by \be c= 3-{6 \over |k|+2},~~~ \hat{c}=2 -{4 \over |k|+2} \ee In
superstring theory, we must tensor it with other conformal field
theories so as to
satisfy the condition $\hat {c}_{tot}=10$ for worldsheet
gravitational anomalies to cancel.

An interesting case considered in
\cite{kounnas-lust} is the case where we add two large
(however compact) free super-coordinates $(y,z)$ together with a compact,
superconformal CFT of central charge $\delta \hat{c}= 6+4/(|k|+2)$.
The resulting background is a four dimensional cosmological
background whose metric in Einstein frame is given by \be
ds_E^2=|k|\alpha'( -dt^2+t^2{dx}^2)+(1+t^2)(R_y^2dy^2+R_z^2dz^2).
\ee This is an anisotropic cosmology which at late times however,
and for large $R_y \sim R_z$, asymptotes to an isotropic flat
Friedman-Robertson cosmology.

The cosmological region $t^2=-uv\ge 0$  is non-compact, and when
$R_{y,z}$ are large it has the desired four-dimensional
interpretation. This is so {\it irrespective of how small the level
$k$ is.} In the region $uv>0$ ($t^2<0$), sigma-model time-like
singularities appear at $uv=1$ ($t^2=-1$). As we propose later in
this work, these singularities are resolved at the string level,
since the structure of the space-time manifold is replaced by a
non-singular $T$-fold.

The string partition function depends crucially on the extra $\hat
c=6+4/(|k|+2)$ superconformal system, which is taken to be compact.
In contrast to the four dimensional part defined by $(t, x, y, z)$,
for the internal, Euclidean $\hat c=6+4/(|k|+2)$ system, the naive
six-dimensional interpretation, which is valid for large level
$|k|$ with
curvature corrections of order $1/(|k|\alpha')$, is not valid
 for small values of
$|k|$ \cite{kiritsis-kounnas}. For example, for $|k|=2$, the system can be
taken to be a {\it  seven}-dimensional torus. In general, small $|k|$
implies that the generalized curvatures (i.e. including dilaton gradients
etcetera)
  are large and the moduli/radii
are small. We remind the reader of the example of the $SU(2)_{k=1}$
Wess-Zumino-Witten
model which is equivalent to a (one-dimensional)
compact boson at self dual radius.
For large level $|k|$, however, the sigma model manifold is a large
 three-dimensional sphere with NSNS 3-form flux.

\section{The Euclidean continuation}
\label{euclidean} Let us consider the region $1- uv\ge 0$ of the
two-dimensional cosmology, and set
$u = -T+X, v=T+X$. We can rotate to Euclidean signature by setting
$T \to -iT_E$. The Euclidean continuation is a disk of unit
coordinate radius parameterized by $ Z=X+iT_E, \bar{Z}=X-iT_E$ such
that $|Z|^2\le 1$. The metric (\ref{metric}) becomes \be
ds^2=|k|\alpha' {dZd\bar{Z} \over 1- Z\bar{Z}}=|k|\alpha'
{d\rho^2+\rho^2d\phi^2 \over 1-\rho^2}\label{metriceucl} \ee and the
dilaton \be e^{2\Phi}={e^{2\Phi_0} \over
1-Z\bar{Z}}={e^{2\Phi_0}\over 1-\rho^2}, \ee where we have also set
$Z=\rho e^{i\phi}$ with $0\le \rho \le 1$. The singularity becomes the
boundary circle $\rho=1$.

The radial distance of the center to the boundary of the disk is
finite, but the circumference of the boundary circle at $\rho=1$ is
infinite. Geometrically the space looks like a bell. This Euclidean
background corresponds to a well defined worldsheet conformal field
theory based on an $SU(2)/U(1)$ gauged WZW model at level $|k|$.

{From} the point of view of the WZW worldsheet theory, the Euclidean
continuation can be understood as a double analytic continuation as
follows. We parameterize the $SL(2,R)$ group manifold as in equation
(\ref{parameter}). Let us also set $a=\tilde{X}-\tilde{T},
b=\tilde{X}+\tilde{T}$ so that the group element becomes \be g=
\begin{pmatrix}
 &\tilde{X}-\tilde{T}  & X-T    \\
                & -X-T  & \tilde{X}+\tilde{T}
\end{pmatrix}
\ee with \be \tilde{X}^2+X^2-\tilde{T}^2-T^2=1. \ee This
parameterization shows that the $SL(2,R)$ group manifold is a
3-dimensional hyperboloid. Then it is clear that upon the double
analytic continuation $T\to -iT_E, \tilde{T}\to -i\tilde{T}_E$ the
group element becomes the following $SU(2)$ matrix \be g =
\begin{pmatrix}
 & W & Z    \\
                & -\bar{Z}  & \bar{W}
\end{pmatrix}
\ee with $W\bar{W}+Z\bar{Z}=1$. After the analytic continuation we also have
that $a\to W=\tilde{X}+i\tilde{T}_E, b\to
\bar{W}=\tilde{X}-i\tilde{T}_E$.

A useful parameterization of the $SU(2)$ group manifold for our
purposes is  \be W=\cos{{\theta}}e^{i\chi}, \, \,
Z=\sin{{\theta}}e^{i\phi} \label{parameterangles} \ee and the metric
on $S^3$ in these coordinates becomes \be ds^2=d\theta^2
+\sin^2\theta d\phi^2 + \cos^2\theta d\chi^2. \ee The ranges of the
angles are as follows $0\le \theta \le \pi, 0\le \chi,\phi,\le
2\pi$.

The original global $SL(2,R)\times SL(2,R)$ symmetry naturally
continues to the $SU(2)\times SU(2)$ global symmetry of the
resulting $SU(2)$ WZW model. The non-compact $U(1)$ axial symmetry
subgroup that we gauge continues to a compact $U(1)$ subgroup
generated by \be \delta g =i\epsilon
\begin{pmatrix}
&1 & 0 \\
           &0 & -1
\end{pmatrix}
g+i\epsilon g
\begin{pmatrix}
&1 & 0 \\
           &0 & -1
\end{pmatrix},
\ee which amounts to the following infinitesimal transformations
$\delta W = 2i\epsilon W$, $\delta \bar{W} = -2i\epsilon \bar{W}$
and $\delta Z= \delta \bar{Z}=0$. In the parameterization
(\ref{parameterangles}), the $U(1)$ symmetry corresponds to shifts
of the angle $\chi$. Gauging this symmetry results in the
$SU(2)/U(1)$ coset model. In the Euclidean set-up, we take the level
$|k|$ to be an integer for the WZW model to be well-defined.

After the analytic continuation described, we end up with the action
(see e.g. \cite{Maldacena:2001ky} for a review): \bea \nn &&
S={|k|\over 2\pi}\int d^2z
\partial\theta\bar{\partial}\theta+
\tan^2\theta\partial\phi\bar{\partial}\phi
 \\
\nn && +\cos^2\theta(\partial\chi+\tan^2\theta\partial\phi+A_z)
(\bar{\partial}\chi-\tan^2\theta\bar{\partial}\phi+A_{\bar z}).\\
\label{actioneucl} \eea In the Euclidean theory the gauge freedom
can be fixed by setting the imaginary part of $W$ (equivalently the
angle $\chi$) to zero. The equations of motion for the gauge field
can then be used to integrate the gauge field out. This amounts to
setting the last term in (\ref{actioneucl}) to zero and producing a
dilaton $e^{2\Phi}=e^{2\Phi_0}/\cos^2\theta$. We end up with a sigma
model action with metric \be ds^2=|k|\alpha' (d\theta^2 +
\tan^2\theta d\phi^2) \label{metriceucl1} \ee which is equivalent to
the metric
(\ref{metriceucl}) after the coordinate transformation $Z=\sin\theta
e^{i\phi}$.

The curvature singularity occurs at $\theta=\pi/2$. The procedure of
fixing the gauge $\chi=0$ and using the equations of motion to
integrate the gauge field out is not valid near $\theta=\pi/2$,
since it results into singular field configurations on the
worldsheet. However, the full action (\ref{actioneucl}) is perfectly
well behaved at $\theta=\pi/2$. To see this, we expand the
Lagrangian in ({\ref{actioneucl}) around $\theta=\pi/2$. Setting
$\theta=\pi/2-\tilde{\theta}$, we obtain that \be S={-|k|\over
2\pi}\int d^2z \phi F_{z\bar z} + O(\tilde{\theta}^2)
\label{actioneucl1}, \ee where we expressed the action in terms of
manifestly gauge invariant degrees of freedom. The leading term in
this expansion describes a simple topological theory, which shows
that an alternative, non-geometric description of the theory can be
given including the region near $\theta=\pi/2$. We return to this
point later on.

From the form of the action near $\theta=\pi/2$, we also learn that
the $U(1)$ symmetry corresponding to shifts of the angle $\phi$ is
quantum mechanically broken to a discrete symmetry
${Z}_{|k|}$. This is because compact worldsheets can support
gauge field configurations for which $\int F_{z \bar z}=2\pi i n$,
with $n$ an integer, and such configurations must be summed over in
the full path integral. It is clear then that the path integral is
only invariant under discrete shifts of the angle $\phi$: $\delta
\phi= 2\pi m/|k|$.

This breaking of the classical $U(1)$ symmetry to ${Z}_{|k|}$
is in accordance with the algebraic description of the $SU(2)/U(1)$
coset in terms of a system of ${Z}_{|k|}$ parafermionic
currents $\psi_{\pm l}(z)$, $l=0,1\dots |k|-1$ [with $\psi_0=1$,
$\psi_l^{\dagger}\equiv \psi_{-l}=\psi_{k-l}$], of conformal weights
$h_l=l(|k|-l)/|k|$ (see also \cite{Bardacki:1990wj}). These satisfy the OPE relations
\begin{eqnarray}
\psi_l(z)\psi_{l'}(0)&=&c_{ll'}z^{-2ll'/|k|}(\psi_{l+l'}(0)+\dots)
\nonumber \\
\psi_l(z)\psi_{l}^{\dagger}(0)&=&z^{-2h_l}(1+ 2h_lz^2 T(0)/c +
\dots)
\end{eqnarray}
which are invariant under the ${Z}_{|k|}$ global symmetry: $\psi_l
\to e^{2\pi i l/k}\psi_l$. Here $T$ is the energy momentum tensor of
the parafermions, $c$ the central charge (which is the
same as the central charge of the coset model)
and the coefficients $c_{ll'}$ are the parafermionic
fusion constants  \cite{Fateev:1985mm}.
In the infinite level
$|k|$ limit, the conformal weights of the parafermion fields become
integers. In this limit the sigma model metric is flat, and we recover the full
rotational invariance of flat space
\cite{Kounnas:1993ix}.
The system can be also generalized to an $N=2$ superconformal system by
tensoring the $Z_{|k|}$ parafermions with a free compact boson
as described
in \cite{Gepner:1987qi}.

Finally we can check that the central charge remains the same after
the analytic continuation. Indeed, it is the very fact that the central
charge of the conformal field theory is smaller than the central
charge corresponding to two macroscopic flat dimensions that codes the de
Sitter nature of the two-dimensional cosmology.

\section{The cosmological T-fold}
\label{tfold}
\subsection*{The parafermionic T-fold}
It is interesting to take a closer look at the geometry that we
associate to the parafermionic model $SU(2)_{|k|}/U(1)$. As we
already discussed, we describe it in terms of a metric and dilaton
profile:
\begin{eqnarray}
ds^2 &=& |k| \alpha' (d \theta^2 + \tan^2 \theta d \phi^2)
\nonumber \\
e^{\Phi} &=& \frac{ e^{\Phi_0} }{\cos \theta },
\label{geom}
\end{eqnarray}
where $\phi \sim \phi+ 2 \pi$ and $\theta$ takes values in the
interval $[0,\pi/2]$. This description breaks down near $\theta =
\pi/2$. Nevertheless, the parafermionic conformal field theory is
perfectly well-behaved, and we can wonder whether there is a more
appropriate, almost-geometrical description. We argue that such a
description exists in terms of a T-fold.

To obtain it, we perform a T-duality along the angular direction
$\phi$ on the geometry described
above:
\begin{eqnarray}
ds^2 &=& |k| \alpha ' d \theta^2 + \frac{\alpha '}{ |k|} \cot^2 \theta d
{\tilde{\phi}}^2
\nonumber \\
 e^{\Phi} &=& \frac{ e^{\tilde{\Phi_0}} }{\sin \theta }. \label{dual}
\end{eqnarray}
By changing variables $\tilde{\theta} = \pi/2 - \theta$, we see that
this is equivalent to:
\begin{eqnarray}
ds^2 &=& |k| \alpha ' d \tilde{\theta}^2 + \frac{\alpha '}{|k|} \tan^2
\tilde{\theta} d \tilde{\phi}^2
\nonumber \\
e^{\Phi} &=& \frac{ e^{\tilde{\Phi_0}} }{\cos \tilde{\theta} }.
\end{eqnarray}
This description is therefore at weak curvature (apart from an orbifold--like
singularity)
and weak coupling
near $\theta=\pi/2$. Moreover, we can identify it as a $Z_{|k|}$
orbifold of a vectorially (or axially) gauged $SU(2)/U(1)$ coset.
Indeed, it is true for the parafermionic theory that the T-dual and
the $Z_{|k|}$ orbifold give two models with identical spectrum due to
the coset character identity $\chi_{j,m}=\chi_{j,-m}$ (see e.g.
\cite{DiFrancesco:1997nk}\cite{Maldacena:2001ky} for reviews).

We now use these facts to give an almost geometrical description of
the parafermionic theory, in terms of a T-fold
\cite{Hellerman:2002ax}\cite{Dabholkar:2002sy}. We use the
description in terms of the first geometry (\ref{geom}) near
$\theta=0$. We cut it just past $\theta=\pi/4$, where the radius of
the circle is $\sqrt{|k| \alpha'}$. We glue it to the T-dual geometry
which we consider near $\tilde{\theta}=0$, or $\theta=\pi/2$, and
which we cut just past $\tilde{\theta}=\pi/4$, where we have radius
$\sqrt{\alpha'/|k|}$. We glue the circles (and their environments)
using the T-duality transformation described above. In the gluing
process, it is crucial to realize that we glue a patch with a
direction of increasing radius to a T-dual patch which in the same
direction has decreasing radius. That gives us the parafermionic
T-fold. The associated partition function is (see
e.g. \cite{Maldacena:2001ky}
for a review) :
\begin{eqnarray}
Z &=& \sum_{j,m} \chi_{j,m} (\tau) \chi_{j,m} (\bar{\tau}).
\end{eqnarray}

One aspect of the model that is rendered manifest by the T-fold
description is the breaking of the $U(1)$ rotation symmetry to a
discrete $Z_{|k|}$ symmetry, due to the $Z_{|k|}$ orbifolding. This is
consistent with our previous discussion of the breaking due to
worldsheet instantons. The T-fold yields an almost-geometrical
picture of the symmetry
 breaking. The T-fold description is indeed everywhere regular
modulo a benign orbifold singularity.

\subsection*{The cosmological T-fold}
In the case of the two-dimensional cosmology as well, we can obtain
a regular T-fold description of the target space of the conformal
field theory. We recall that under T-duality (the metric can be obtained
by analytically continuing the metric
 (\ref{dual}) in the direction $\tilde{\phi}$), the light-cone and the
singularities get interchanged.
 Consider the cosmology, and cut it
at a hyperbola at radius $\sqrt{|k| \alpha'}$, in between the
light-cone and the time-like singularities in the Penrose diagram
(see the upper part of figure \ref{cosmologyanddisctfold}). Consider
then its T-dual, and cut it along a similar line. Glue the two parts
of the T-dual cosmologies along these cuts to obtain the T-fold
cosmology.
%[METRICS and explicit GLUING in this region needs to be added HERE.]
The description we obtain is particularly nice as we no longer need
a microscopic origin of a would-be source associated to the
time-like singularities, nor do we need to define boundary
conditions associated to them. There is no singularity in,
nor is there a boundary to the T-fold cosmology.
 Indeed, the almost-geometrical
description is very much like $dS_2$, which we can think of as a
hyperboloid embedded in three-dimensional space. The difference is
that the T-fold cosmology has two patches glued together via a
T-duality transformation (instead of an ordinary coordinate transformation in
the case of two-dimensional de Sitter space).

In figure \ref{cosmologyanddisctfold} we show how the T-fold
description of the parafermions and the two-dimensional cosmology
continue into one another after analytic continuation.
\begin{figure}
\centering
\includegraphics[width=10cm]{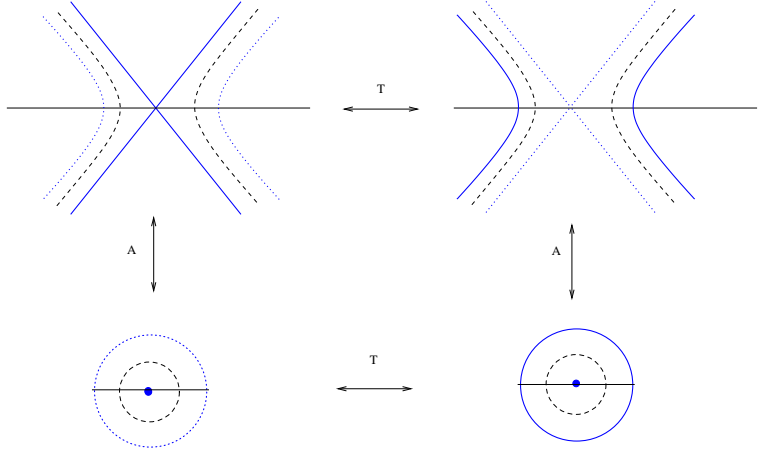}
\caption{The analytic continuation of the T-fold. The upper part of the
  diagram shows the two T-dual descriptions of the cosmology in which
the horizon and the (apparent) singularities are exchanged (in bold blue).
The (striped black) cut along which they are glued is indicated,
as well as the (thin black) line along which the cosmology is cut to obtain
a space-like slice (see later). Analytic continuation then gives rise to the lower
part of the figure, in which we have sketched the T-fold description of the
parafermionic conformal field theory. In bold blue we have the center and
the boundary of the disk, and (in black stripes) the T-dual circles along
which we glue.}
\label{cosmologyanddisctfold}
\end{figure}

\section{Defining the wave-function of the universe}
\label{wavefunction} Later on, we will consider string theory
backgrounds which are product models and in which one factor
consists of the two-dimensional cosmology discussed in sections
\ref{cosmo}, \ref{euclidean} and \ref{tfold}. For these models, we wish
to define a wave-function of the universe in string theory following ideas of
\cite{Hartle:1983ai} which define a wave-function of $dS_n$
universes within a field theoretic context.

We consider a time-reversal symmetric space-like slice of the
cosmology, within the boundaries of the (seeming) singularities.
See figure \ref{cosmologyanddisc}. This is the slice $T=0$. In the
past of the space-like slice, we glue half of the target space of an
 $SU(2)/U(1)$ coset conformal
field theory -- a half disk. By the analytic continuation discussed
in the previous section, this gluing is continuous in the
backgrounds fields, and moreover in the exact conformal field theory
description.
 \begin{figure}
\centering
\includegraphics[width=6.5cm]{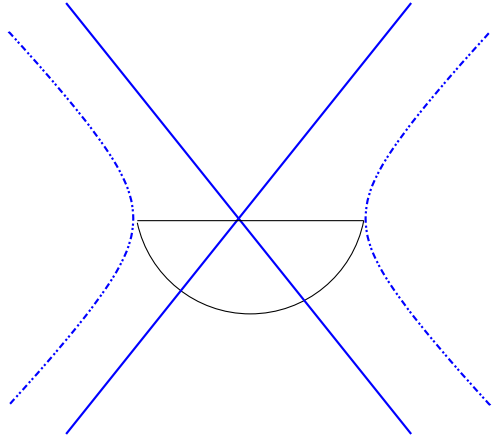}
\caption{The continuous gluing of the half-disk into the cosmology, when
cutting the cosmology along a space-like slice, and analytically continuing.
The figure should be viewed as a simplified version of the previous T-fold
picture.}
\label{cosmologyanddisc}
\end{figure}

A crucial feature of the proposal of \cite{Hartle:1983ai} for the
definition of the wave-function of the universe is that the corresponding
Euclidean space
is without boundary. In our set-up as well, the Euclidean conformal
field theory has a target with no boundary. It is important in this
respect that we have obtained an almost-geometric description of the
parafermionic conformal field theory\footnote{A
  traditional description of the target space as a disk, which is
  singular,would lead to the faulty conclusion that the target space has a
boundary.}.
It is intuitively clear from the T-fold description
given in the previous section that the parafermionic theory does not
have a boundary\footnote{Since a T-fold is non-geometric, one
  needs to define the concept of boundary precisely. We believe that a
  reasonable definition will match our intuition.}. When we cut the
  Euclidean T-fold into half, it is clear (from figure
  \ref{cosmologyanddisctfold}) that we can glue the boundary of that
  half-T-fold into the initial surface of the cosmological T-fold.

Thus we have determined the precise gluing of T-folds necessary in
order to define a wave-function depending on initial data. We
now define the wave-function of the universe by performing a
``half T-fold'' Euclidean
path integral over {\it all} target space
fields with specified values on the
boundary:
\begin{eqnarray}
\Psi [ h_\partial ,\phi_\partial,\dots ] &=& \int [dg]
[d\phi] \dots e^{-S(g, \phi, \dots)}~, \label{path}
\end{eqnarray}
where the path integral is such that the metric, the dilaton and all
other space-time fields satisfy $g=h_\partial, \phi=\phi_\partial, \dots$
on the boundary of the half T-fold that we glued into the cosmological
solution.
The path integral
above can in principle be performed off-shell, in a second-quantized
string field theory context, where we may also express it as an integral
over a single string field $\Phi$. (See e.g. \cite{Polchinski:1998rq} for a
concise review).

Let us be more specific. The initial space-like slice of the T-fold
cosmology has two patches. On each patch, we define a boundary
metric $h^1$ and $h^2$, and similarly for other fields. The boundary
metrics satisfy the condition that on the overlap of the patches,
they match up to a T-duality transformation, symbolically:
$h^1_{\partial^1 \cap \partial^2} = T (h^2_{\partial^1 \cap
\partial^2})$. This is the way in which we can specify boundary data
precisely. In the following, we do not emphasize this important part
of the definition of the path integral further, not to clutter the
formulas.

In principle, a T-fold path integral can be computed as
follows. Consider again the two patches.
Each patch has a non-singular geometric
description. Over each patch the path integral reduces to an
ordinary field theory path integral, and can be performed in the usual way
giving rise to a functional of boundary data. The full path integral
can be obtained by integrating the two functionals together over
data that belong to the common boundary of the two patches.
Since at the common
boundary of the patches their fields are related by a
T-duality transformation,
to do the final integral we
would need to perform a T-duality transformation on one of the two
functionals.

We remark here that this particular feature of definitions
of path integrals over T-folds with boundaries is generic. The above
description is easily extended to a generic description of T-fold
boundary data. Although we do not need a general prescription in this
paper,
we believe it would
be interesting to develop the path integral formalism for T-folds with
boundary further.

The prescription for the wave-function of the universe we
outlined above should have an
analogue, via the relation between string oscillators and
the target space fields, to a first quantized prescription. Notice
that the initial-time data allow multi closed string configurations. Summing over
histories that lead to them, would allow worldsheets with boundaries (and other
topology
features),
including disconnected ones.
The wave-function would take the form
\begin{eqnarray}
\Psi [ X_\partial (\sigma, \tau) ] &=& \sum_{\rm topologies} \int  [dX]
e^{-S[X(\sigma,\tau)]},
\end{eqnarray}
where the worldsheet path integrals are performed over string
configurations $X(\sigma, \tau)$ that satisfy a specified boundary
condition at given values of the zero-modes of the string
configuration, i.e. at a given position of the target space. The
equivalence of these descriptions is far from obvious, but it is
made plausible by the fact that for two-dimensional string
worldsheets, the first quantized description automatically comes
with a prescription for the proper weighting of interaction
vertices. The initial-time closed string configurations could be
specified in terms of macroscopic loop operators discussed for
example in \cite{Moore:1991ag}. The first quantized prescription
considers fluctuations around a given background. A full second
quantized prescription also integrates over backgrounds as in
general relativity \cite{Hartle:1983ai}.

The wave-function so defined is hard to compute, although it may be
obtained presumably for very particular boundary conditions. An
example would be boundary conditions that are fixed by taking a
$Z_2$ orbifold that folds over the disk onto itself -- in that case,
one may be able to compute the value of the wave-function for a
particular argument. In order to understand better some global
properties, we again follow \cite{Hartle:1983ai}  and concentrate on
calculating the norm of the wave-function.

\section{The norm of the wave-function}
\label{norm} The norm of the wave-function is easier to compute. It
is given by the following calculation:
\begin{eqnarray}
||\Psi||^2 &=& \int [d\Phi_\partial] \int_{ {\rm half~ T-fold}} [d\Phi]
e^{-S(\Phi)}  \times
\nonumber \\
& & \int_{{\rm conj \,\, half \,\, T-fold}} [d\Phi]  e^{-S(\Phi)}
\nonumber \\
&=& \int_{{\rm T-fold}} [d \Phi] e^{- S(\Phi)},\label{pathfull}
\end{eqnarray}
where we have expressed it as a string field theory path integral in
terms of a string field $\Phi$. The final integral is an integral
over all possible string field configurations on the Euclidean
T-fold. No boundary conditions need to be specified.

We can do this calculation by considering the fluctuations
around an on-shell closed string background, in a first quantized
formalism:
\begin{eqnarray}
||\Psi||^2 &=& \sum_{\rm topologies} \int  [dX] e^{-S[X(\sigma, \tau)]}
\end{eqnarray}
where $X(\sigma,\tau)$ is any mapping from the string worldsheet
into the target space. The sum is over all closed worldsheet
topologies, and includes a sum over disconnected diagrams. In fact
it is equal to the following exponential of a sum of connected
diagrams:
\begin{eqnarray}
||\Psi||^2 &=& \exp ( Z_{total} ),
\end{eqnarray}
where the function $Z_{total}$ is the total string theory partition
function, which is defined as a sum over Euclidean worldsheet
topologies:
\begin{eqnarray}
Z_{total} &=& \frac{1}{g_s^2} Z_{S^2} + Z_{T^2} + g_s^2 Z_{genus=2}
+ \sum_{g=3}^{\infty} g_s^{2g-2} Z_{genus=g}.
\end{eqnarray}
Therefore, to evaluate the norm of the wave-function perturbatively, we need to
evaluate the partition function for string theory on the Riemann
surfaces of genus $0,1,2,\dots$ and add their contributions with the
appropriate power of the string coupling constant. The first
contribution is akin to the tree level contribution in ordinary
gravity, the second to the one-loop contribution, etc.

\section{Thermal nature of the wave-function}

A natural way to perform the Euclidean path integral
in equation (\ref{path})
over half the
space is as follows. The origin $X=0$ in one T-fold patch (and
similarly for the other), divides the $T=0$ slice into two halves:
the left half corresponding to $X<0$ and the right part
corresponding to $X>0$. We denote the boundary data on $X<0$ by
$\phi_L$ and on $X>0$ by $\phi_R$. See figure
\ref{thermal}. By dividing the space into
angular wedges spanning an overall angle equal to
$\pi$, we can evaluate the path integral in terms of the generator
of angular rotations.
This generator is given by the analytic continuation of
$iH_{\omega}$, where $H_{\omega}=i\partial_{\omega}$ is the Hamiltonian conjugate to
``Rindler'' time in the region $uv>0$ of the Lorentzian cosmology.
Indeed in this region, we may set $u=\rho e^{-\omega}, ~ v=\rho
e^{\omega}$, with the string frame metric and dilaton given by 
$$ ds^2 = |k|\alpha'
{d\rho^2 - \rho^2 d\omega^2 \over 1-\rho^2}$$
\be
e^{2\Phi}= {e^{2\Phi_0} \over 1-\rho^2}. \label{rindler2}
\ee
In this patch, the background metric is static, invariant under time translations, 
and the dilaton field is space-like.
 Rotating to
Euclidean signature amounts to setting $\omega=-i\phi$. So Rindler time
translations correspond to angular rotations in the Euclidean space.
As we have already discussed, only discrete angular rotations are true
symmetries of the string theory background.

 \begin{figure}
\centering
\includegraphics[width=10cm]{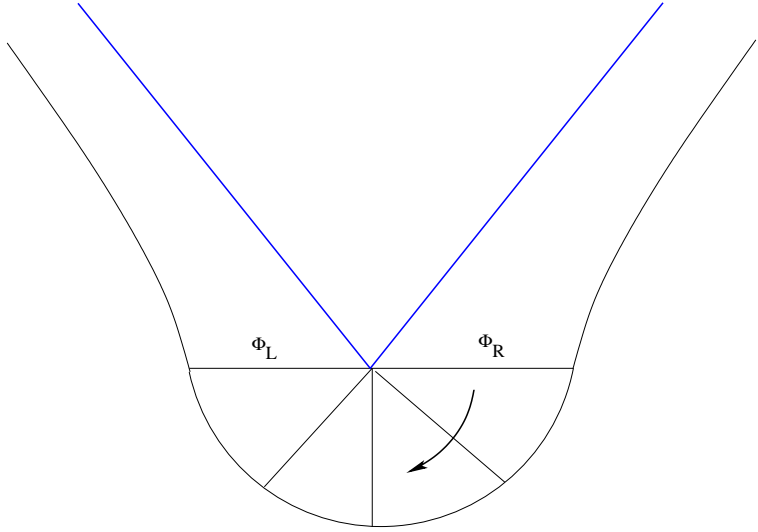}
\caption{The thermal interpretation of the wave-function is obtained
by thinking of the path integral as being performed along angular wedges from
an initial (right) to a final (left) configuration.}
\label{thermal}
\end{figure}

The boundary data can then be viewed as specifying initial {\em and}
final conditions for the path integral evolution. This is clearly
reviewed for the case of flat Rindler space and black hole spaces
in \cite{Bigatti:1999dp}. In
particular, the path integral measures the overlap between the data
on the right $\phi_R$, evolved for a Euclidean time $\pi$, and the
data specified on the left $\phi_L$ (see figure
\ref{thermal}), and it can be written as an amplitude \be
\Psi(\phi_L, \phi_R)=\langle \phi_L| e^{-\pi H_{\omega}}|
\phi_R\rangle. \label{rindler} \ee
If we integrate over $\phi_L$ we obtain a thermal density matrix appropriate for the
Rindler observer \cite{Bigatti:1999dp}
\be
\int [d\phi_L]\Psi(\phi_L, \phi_R)\Psi^*(\phi_L, \phi_R^{\prime})=\langle
\phi_R^{\prime}| e^{-2\pi H_{\omega}}|
\phi_R\rangle,
\ee
with dimensionless temperature $T_{\omega}=1/2\pi$.
The norm of the
wave-function is given by the trace
\be
||\Psi||^2={\rm Tr} e^{-2\pi H_{\omega}}
\ee
and so it can be interpreted as a thermal space-time partition function.
The genus-1 string contribution
is a thermal one-loop amplitude.

In the full Euclidean path integral equation (\ref{pathfull}), the
contributions of the fermionic fields are positive. To perform the
full path integral over the whole T-fold, we divide it into angular
wedges spanning an angle equal to  $2\pi$. Since the space has no
non-contractible cycles, the space-time fermionic fields have to be
taken anti-periodic in the angular variable and they contribute
positively to the path integral.

The T-fold patches are glued along the hyperbola $\rho=1/\sqrt{2}$
(see section \ref{tfold}). Near this region, the curvature is low
for large enough level $|k|$. Thus for large level $|k|$, we may use
the metric equation (\ref{rindler2}) to conclude that observers
moving near the region $\rho \sim 1/\sqrt{2}$ measure a proper
temperature
%of
%order
$T \sim 1/(2\pi \sqrt{|k|\alpha'})$. In the cosmological region $uv \le 0$,
there is also an effective temperature of the same order as a result of 
particle production \cite{cornalba}\cite{Toumbas:2004fe}.
 For small level $|k|$, we need
a string calculation to deduce the proper temperature of the system.

\section{Specific Examples}
\label{models}

As we discussed above in order to derive the wave-function of the
cosmology, we need to compute the total string partition function
for the corresponding Euclidean background. When
conformal field theories are compact, the genus-zero contribution to
the total string partition function vanishes. This is because the
spherical string partition function is divided by the infinite volume of the
conformal Killing group. This fact is a first important difference
with the calculation in general relativity where the classical
contribution is non-zero. In perturbative string theory
the leading contribution is the genus-1 amplitude.

The Euclidean examples we shall describe here in detail belong to
the family of $\hat{c}=10$ superconformal, compact systems. In order
for them to admit a Lorentzian continuation to a cosmological
(time-dependent) background, space-time supersymmetry must be
broken. Moreover, the models must be free of tachyons.
The presence of tachyonic modes would indicate that the system
undergoes a phase transition.
The only known examples
with the above properties are of the form
\begin{equation}
{SU(2)_{|k|} \over U(1)}~~ \times~~ K \longleftrightarrow ~~{SL(2,
R)_{-|k|} \over U(1)}~~ \times~~ K~,
\end{equation}
where we indicated the analytic continuation from the Euclidean to the
Lorentzian space-time.
The level $|k|$ can be taken to be
small.

As we already discussed, the relevant genus-1 string amplitude has to be thermal.
The total superstring model has transverse central charge equal to $c=12$ (or
$\hat{c}=8$). As a consequence, it has a Hagedorn transition at the fermionic
 radius $R_H = \sqrt{2 \alpha'}$. In order for the genus-1 string amplitude to be finite,
the physical temperature of the model has to be below the Hagedorn temperature: 
$T < T_H=1/(2\pi\sqrt{2 \alpha'})$.
Let us give an argument that this can be realized for any $|k|\ge 2$. 
In writing the norm of the wave-function as a thermal space-time partition function,
the role of the Hamiltonian is taken up by
 the generator of rotations on the disk. For the
superconformal $SU(2)/U(1)$ model,
the corresponding $U(1)$ current is 
at level $|k|+2$ \cite{Gepner:1987qi}. 
Thus we expect the physical temperature of
 the model to be set by the radius associated to this isometry generator,
 namely $\sqrt{(|k|+2) \alpha'}$. The corresponding temperature is given by
\be
T= {1 \over 2\pi\sqrt{(|k|+2) \alpha'}} \label{temperature}
\ee
and is below the Hagedorn temperature for any positive (integer) level $|k|$. 
We will find that this temperature arises naturally in a level $|k|=2$ model below.
At level $|k|=0$, where the minimal model has zero central
 charge, and consists only of the identity operator (and state), the cosmology
 disappears. When we reach the Hagedorn transition, the cosmology becomes so
 highly curved that it is no longer present in the string theory background.
%The temperature grows as the curvature grows, but the number of states
%diminishes at the same time, until only one state is left.

\subsection{Compact models}
In a first class of specific examples that we will discuss in this section,
we choose the level $|k|=2$, and
we take the internal conformal field theory K to be:
\begin{equation}
K=T^2 \times \prod_{i=1,\dots,7}{SU(2)_{k_i} \over U(1)}
\end{equation}
where all $k_i$'s are taken equal to $2$, so that $\hat{c}_K=9$ (representing
the central charge equivalent of nine flat directions). 
In the sequel, we set $\alpha'=1$.

For this choice, the supersymmetric characters of the whole system
are defined in terms of eight level $k=2$ parafermionic systems
(which are nothing but eight real fermions $\psi_i$ and eight bosons
$\phi_i$ compactified at the self-dual radius $R=1$), and also
a complex fermion $\Psi_T$ and a complex boson $\Phi_T$ for the torus
$T^2$. The $N=2$ superconformal operators $T_F, J$ are:

$$
T_F~=~\sum_{i= 0,1,\dots 7}\psi_i~e^{i\sqrt{2}\phi_i }~ +i\Psi_{T}
\partial \Phi_{T}
$$
\begin{equation}
J={i\over \sqrt{2}}\sum_{i= 0,1, \dots 7}\pd\phi_i+ \Psi_T
\bar{\Psi}_T.
\end{equation}

It is convenient to pair the $(0,1),~(2,3),~(4,5),~(6,7)$ systems respectively
in order to obtain
four copies of $\hat{c}=2$ systems. For the first copy we define the
bosons ${\cal H}_0,~{\cal H}_1$ at radius $R=\sqrt{2}$  (or  $R={1\over \sqrt{2}}$ ,
fermionic  $T$-dual points):

\begin{equation}
\phi_0={1\over \sqrt{2}}~({\cal H}_0+{\cal H}_1),~~ \phi_1=~{1\over
\sqrt{2}}~({\cal H}_0-{\cal H}_1)~~~
\end{equation}
and similarly for the others. Then the currents are given by
$$
T_F=\sum_{i=0,2,4,6}\psi_i  e^{i({\cal H}_i+{\cal H}_{i+1})}
+\psi_{i+1}~e^{i({\cal H}_i-{\cal H}_{i+1})} +i\partial \Phi_T e^{i{\cal H}_T} $$
\begin{equation}
  J=i\pd {\cal H}_0+i\pd {\cal H}_2+i\pd {\cal H}_4+i\pd {\cal H}_6+ i\pd {\cal H}_T,
\end{equation}
where $i\pd {\cal H}_T= \Psi_T \bar{\Psi}_T$, which is also defined at the
fermionic point. Observe that the $N=2$ current is given in terms of
${\cal H}_0$, ${\cal H}_2$, ${\cal H}_4$, ${\cal H}_6$ and ${\cal H}_T$ {\em only} and is normalized
correctly for a system with ${\hat c}=10$.

The $N=2$ left-moving characters of a particular $\hat c =2$ system (e.g
of the one containing ${\cal H}_0$), are expressed in terms of the usual level-2
$\Theta$-functions \footnote{Our convention for the level-2
$\Theta$ function is $\Theta \left[^{\gamma}_{\delta}\right]=\sum
_{n}~e^{i\pi \tau(n+{\gamma\over 2})^2+2i\pi (n+{\gamma\over
2})({\delta\over 2})}$.}:
 \be {1\over \eta(\tau)^3}~
\Theta_{\psi_0}\left[^{\gamma_0}_{\delta_0}\right]^{1\over 2}
\Theta_{{\cal H}_0}\left[^{\alpha +2H_0} _{\beta+2G_0}\right]
\Theta_{{\cal H}_1}\left[^{\gamma_0 -2H_0}_{\delta_0 -2G_0}\right]
\Theta_{\psi_1}\left[^{3\gamma_0 -4H_0} _{3\delta_0
-4G_0}\right]^{1\over 2} \ee where the arguments
$(\gamma_0,\delta_0)$ and $(4H_0,4G_0)$ are integers.  The later
exemplifies the chiral $Z_{k+2}$-symmetry of the superconformal
parafermionic characters ($k+2=4$ in our case). Similar expressions
are obtained for the other three $\hat c=2$ parafermionic systems by
replacing $(\gamma_0,\delta_0)$ with $(\gamma_i,\delta_i)$,  and
$(H_0,G_0)$ by $(H_i,G_i), ~i=1,2,3$.

The global existence of the
$N=2$ superconformal world-sheet symmetry and thus the existence of the left space-time
supersymmetry imply \be (H_T,G_T) +\sum_{i=0,1,2,3}~ (H_i,G_i)= \epsilon ~{\rm mod~2}~, \ee and
similarly for the right supersymmetry. 
The arguments $(\epsilon, \bar\epsilon)$ define the chirality of the
space-time spinors.
A simple choice is to set
$(H_T, G_T)=(0,0)$, $(H_i,G_i)=(H,G)$ for $i=0,1$ and $(H_i,G_i)=(-H,-G)$ for $i= 2,3$ (and
similarly for the right arguments). Then if $(\epsilon,
\bar\epsilon)=(1,1)$, space-time supersymmetry is broken. For the
other choices there is some amount of supersymmetry preserved. In
this class of models the only remaining possibility consistent with
the global $N=2$ super-parameterization consists of shifts on $\Phi_T$ of
the $T^2$ torus.

\subsubsection*{$Z_4$ orbifold models}

Using the chiral $Z_4$ symmetry defined above and its subgroups,
we can obtain four classes of models: \be (H,G)={M\over
4}(h,g),~~~{\rm where} ~~(h,g)=~{\rm integers}, ~~~~M=1,2,3,4. \label{orbifold}\ee
 In particular, if we orbifold by
$Z_4$ ($M=1$ or $ M=3$ ), the $\psi_0$ parafermion decouples from the
rest, especially from $\psi_1$. This is clear since in this case,
the arguments of $\Theta_{\psi_0}$ and $\Theta_{\psi_1}$ become
independent.

Initially, the arguments $(\gamma_i,\delta_i)$ are taken to be
identical for the left- and right-moving characters. In this
case the modular invariance of the partition function is manifest.
Indeed, using  the periodicity property of
$\Theta$-functions
\be \left|\Theta\left[^{\gamma
-2\epsilon}_{\delta -2\zeta}\right]\right|=
 \left|\Theta\left[^{\gamma}_{\delta}\right]\right|,
 ~~~~\epsilon,\zeta ~~{\rm integers},
\ee
and orbifolding by $Z_4$, ($M=1$), the genus-1
modular invariant partition function becomes:
$$
Z=\int _F {d\tau d\bar\tau \over ({\rm Im} \tau)^2}~{{\rm Im} \tau \over
\eta(\tau)^{12}\bar\eta(\bar\tau)^{12}}~ {1\over
4}~~\sum_{h,g=0,1,2,3}\Gamma_{2,2}
$$
$$
\times{1\over 2} \sum_{\gamma_0,\delta_0}
 \left|\Theta\left[^{\gamma_0}_{\delta_0}\right]\right|~
 \left|\Theta\left[^{\gamma_0-h} _{\delta_0-g}\right]\right|~
 \left|\Theta\left[^{\gamma_0-{h\over2} }_{\delta_0-{g\over2}}\right]\right|^2
$$
$$ \times{1\over 2}
\sum_{\gamma_1,\delta_1}
 \left|\Theta\left[^{\gamma_1}_{\delta_1}\right]\right|~
 \left|\Theta\left[^{\gamma_1-h} _{\delta_1-g}\right]\right|~
 \left|\Theta\left[^{\gamma_1-{h\over2} }_{\delta_1-{g\over2}}\right]\right|^2
$$
$$
\times{1\over 2} \sum_{\gamma_2,\delta_2}
 \left|\Theta\left[^{\gamma_2}_{\delta_2}\right]\right|~
 \left|\Theta\left[^{\gamma_2+h} _{\delta_2+g}\right]\right|~
 \left|\Theta\left[^{\gamma_2+{h\over2} }_{\delta_2+{g\over2}}\right]\right|^2
$$
$$
\times{1\over 2} \sum_{\gamma_3,\delta_3}
 \left|\Theta\left[^{\gamma_3}_{\delta_3}\right]\right|~
 \left|\Theta\left[^{\gamma_3+h} _{\delta_3+g}\right]\right|~
 \left|\Theta\left[^{\gamma_3+{h\over2} }_{\delta_3+{g\over2}}\right]\right|^2
$$
$$
\times ~{1\over 2}\sum_{(\alpha,\beta)}~e^{i\pi(\alpha+\beta
+\epsilon\alpha\beta)}
~~\Theta\left[^{\alpha+{h\over2}}_{\beta+{g\over2}}\right]^2
\Theta\left[^{\alpha-{h\over2}}_{\beta-{g\over2}}\right]^2
$$
\be \times ~{1\over 2} \sum_{(\bar\alpha,\bar\beta)}~
e^{i\pi(\bar\alpha+\bar\beta +\bar\epsilon\bar \alpha\bar\beta)}
~~\bar\Theta\left[^{\bar\alpha
+{h\over2}}_{\bar\beta+{g\over2}}\right]^2\bar\Theta\left[^{\bar\alpha
-{h\over2}}_{\bar\beta-{g\over2}}\right]^2. \label{partition} \ee
The arguments $(\alpha, \beta)$ and $(\bar\alpha, \bar\beta)$ are
those associated to the $N=2$ left- and right-moving supercurrents.
%The arguments $(\epsilon, \bar\epsilon)$ define the chirality of the
%space-time spinors. 
If $(\epsilon, \bar\epsilon)=(1,1)$,
supersymmetry is broken. For all other choices this partition function
is identically zero and there is some amount of supersymmetry
preserved.
To obtain the above result we have used the fact that the contribution of
the superconformal ghosts cancel the oscillator contributions of the
$T^2$ supercoordinates $(\Phi_T, \Psi_T)$. This is the reason for
choosing the $Z_4$ not to act on $\Psi_T$, having set $(H_T, G_T)=(0,0)$. 
The only remnant from the
torus contribution is the $\Gamma_{2,2}$ lattice, which can be
possibly shifted by $(Lh/2,Lg/2)$ with either  $L=0$ or $L=1,2,3$ (as
we will see later).

%This partition function is not yet thermal and it is not
%the relevant one for the cosmological wave-function.
To proceed we need to identify and insert the thermal co-cycle
$S\left[^{q\,,~(\alpha\,+\,\bar\alpha)}_{p,~(\beta+\bar\beta)}
\right]$ associated to the time direction of the cosmology: \be
S\left[^{q,\,~(\alpha+\bar\alpha)}_{p,~(\beta+\bar\beta)} \right]
=e^{i\pi
\left(\,p(\alpha+\bar\alpha)\,+\,q(\beta+\bar\beta)\,\right)}, \ee
where $p$ and $q$ are the lattice charges associated to the
compactified Euclidean time. Here, $F_{\alpha}=(\alpha+\bar\alpha)$ and
$F_{\beta}=(\beta+\bar\beta)$ define the spin of space-time
particles: $F_{\alpha}=1$  modulo $2$ for fermions and $F_{\alpha}=0$
modulo $2$ for bosons. To impose this co-cycle insertion, it is
necessary to rewrite the partition function in a form that 
reveals the charge lattice $(p,q)$ of the Euclidean time
direction.
To this end, it is convenient to separate the partition function into the
``untwisted sector''  $(h,g)=(0,0)$ and `` twisted sectors''
$(h,g)\ne (0,0)$,
\be Z=\int _F {d\tau d\bar\tau \over ({\rm Im} \tau)^2}
%Im \tau\Gamma_{2,2}~{1\over 4}
\left( Z_{{ unt}}+ \sum_{(h,g)\ne
(0,0)} Z_{twist}\left[^h_g \right] \right).\ee

To isolate the relevant $(p,q)$ charge lattice, we use the identity
\be
\left|\Theta\left[^{\gamma}_{\delta}\right] \right|^2  =
 {R\over
\sqrt{{\rm Im}\tau}}~\sum_{(m,n)} ~e^{-\pi R^2{|m+\tau n|^2\over {\rm Im}\tau}}
~e^{i\pi ( m \gamma+ n\delta + mn)} ~,~~~~R^2 ={1\over 2}.
\ee
Although in the above identity the radius is fixed to the fermionic
point $R^2 = 1/2$, we note that
{\em the modular transformation properties are the same for any
$R^2$}, and in particular for the dual-fermionic point with $R^2=2$.

Using the above identity for the conformal block involving the
parafermions $\psi_{0,1}$ and the ${\cal H}_1$ field, we obtain
$$
{1\over 2}\sum_{(\gamma,\delta)}\left|\Theta_{\psi_{0,1}}
\left[^{\gamma}_{\delta}\right] \right|^2 \left|\Theta_{{\cal H}_1}
\left[^{\gamma }_{\delta}\right]\right|^2~=
$$
\be {1\over 2} \sum_{(\gamma,\delta)} ~\sum_{(m_1,n_1),(m_2,n_2)}
{R_1R_2\over {\rm Im}\tau}~e^{-\pi R_1^2{|m_1+\tau n_1|^2\over {\rm Im}\tau}-\pi
R_2^2{|m_2+\tau n_2|^2\over {\rm Im}\tau
}}~e^{i\pi\left((m_1+m_2)\gamma+(n_1+n_2)\delta
+m_1n_1+m_2n_2\right)} \label{block} \ee where
$R_1^2=R_2^2=R^2={1/ 2}$. Since the arguments
$({\gamma},{\delta})$ do not appear elsewhere (in equation (\ref{partition})), summing over them
forces $(m_1+m_2)$ and $(n_1+n_2)$ to be even integers. This
constraint can be solved if we take \be m_1=p_1 +p_2, ~~~m_2= p_1
-p_2,~~~~ n_1=q_1+q_2, ~~n_2= q_1- q_2 \label{constraint} \ee so
that,
$$
{1\over 2}\sum_{(\gamma,\delta)}\left|\Theta_{\psi_{0,1}}
\left[^{\gamma}_{\delta}\right] \right|^2 \left|\Theta_{{\cal H}_1}
\left[^{\gamma }_{\delta}\right]\right|^2~=~~
\Gamma_{1,1}(R_+)~~\Gamma_{1,1}(R_-)~=
$$
\be \left(\sum_{(p_1,q_1)} {{\rm Im}\tau}^{-{1 \over 2}}~R_+~e^{-\pi
R_+^2{|p_1+\tau q_1|^2\over
{\rm Im}\tau}}\right)~\left(~\sum_{(p_2,q_2)}~{{\rm Im}\tau}^{-{1 \over 2}}~
R_-~e^{-\pi R_-^2{|p_2+\tau q_2|^2\over {\rm Im}\tau }}\right) \ee with
$R^2_+=R^2_-=2R^2=1$. Therefore, the partition function of the {\em
bosonic} part of parafermions {\em factorizes in two}
$\Gamma_{1,1}$ lattices, both of them with twice the initial radius
squared.

The charge lattices $(p_1,q_1)$ and $(p_2, q_2)$ are
associated to the $\psi_0$ and
$\psi_1$ parafermions. We can see this as follows. Consider the
left-charge operators which are well defined in the untwisted
sector: \be Q_+= i\oint~dz(\psi_0\psi_1 +
\partial {\cal H}_1), ~~~~Q_-=i\oint~dz(\psi_0\psi_1 - \partial {\cal H}_1) \ee
and similarly for the right-moving ones ($\bar Q_{\pm}$). Then,
$$
(Q+\bar Q)_{+}=m_1+ m_2=2p_1, ~~~~(Q-\bar Q)_+=n_1+n_2=2q_1
$$
\be (Q+\bar Q)_{-}=m_1-m_2=2p_2, ~~~~(Q-\bar Q)_-=n_1-n_2=2q_2 \ee
where we have used the constraint (\ref{constraint}). We identify
the charges $(p_1,q_1)$ as the momenta that enter in the thermal
co-cycle, and associate the lattice to the Euclidean time direction.

Before we proceed further, let us stress the following point. We
started with a diagonal modular invariant combination and with
initial radii $R_1^2=R_2^2=R^2={1/ 2}$. The anti-diagonal choice
implies that the initial values for the radii are at the fermionic
T-dual points, namely  $R_1^2=R_2^2=R^2=2$. Notice that in all we
perform two T-dualities simultaneously so that we remain in the same
type II theory. Thus the conformal block, equation (\ref{block}), can be
replaced with the T-dual one with $R^2=2$.\footnote{In Gepner's
formalism \cite{Gepner:1987qi} the anti-diagonal combination
corresponds to exchanging $(m,\bar m) \to (m, -\bar m)$.} For the
anti-diagonal choice, the radii of the corresponding factorized
lattices are given by $R_+^2=R_-^2=4$ instead of unity for the
diagonal combination.

Thus in the untwisted sector, $Z_{unt}$ has to be replaced with \be
Z_{unt}\longrightarrow
Z^{thermal}_{unt}=\sum_{(p_1,q_1)}~\sum_{(\alpha,\beta),(\bar
\alpha, \bar\beta)} ~Z^{thermal}_{unt}\left[^{q_1,~
\alpha+\bar\alpha}_{p_1, ~\beta+\bar\beta
}\right]~e^{i\pi\left(p_1(\alpha+\bar
\alpha)+q_1(\beta+\bar\beta)\right)}. \ee Performing a similar
factorization for the remaining three copies of $\hat c=2$
superconformal parafermionic blocks, we can write the thermal untwisted
contribution in a compact form: \be Z^{thermal}_{unt}={1\over 4}~
{{\rm Im}\tau\Gamma_{9,9}\over \eta^{12} \bar\eta^{12}}~~\left\{{1\over
4}\sum_{(\alpha,\beta),(\bar\alpha,\bar\beta)}~\Gamma_{1,1}(R_+)\left[
^{{\alpha+\bar\alpha}}_{{\beta+\bar\beta}}\right]~e^{i \pi  (\alpha+
\beta+\epsilon \alpha\beta)}~ \Theta\left[^{\alpha}_{\beta}\right]^4
e^{i \pi (\bar\alpha+\bar\beta +\bar\epsilon\bar\alpha\bar\beta)}
~\bar\Theta\left[^{\bar\alpha}_{\bar\beta}\right]^4 \right\}. \label{untthermal}\ee  The
$\Gamma_{9,9}$ lattice factor is composed of a product of lattices:
the initial $\Gamma_{2,2}$ lattice of the torus $T^2$ with radii $R_y,
R_z$, the $\Gamma_{1,1}$ lattice of the first parafermionic block
at radius $R_-$, and the product of three pairs
$\Gamma_{1,1}(R^i_+)~\Gamma_{1,1}(R^i_-)$ for the other three
parafermionic blocks. The $\Gamma_{1,1}(R_+)$ lattice is the
thermally shifted lattice \be
 ~\Gamma_{1,1}(R_+)\left[
^{{\alpha+\bar\alpha}}_{{\beta+\bar\beta}}\right]~
=~\sum_{(p_1,q_1)} {{\rm Im}\tau}^{-{1 \over 2}}~R_+~e^{-\pi
R_+^2{|p_1+\tau q_1|^2\over {\rm
Im}\tau}}e^{i\pi\left((\alpha+\bar\alpha)p_1+(\beta+\bar\beta)q_1\right)
}. \ee Its coupling with the space-time spin structure breaks
space-time supersymmetry so that both bosons and fermions give
positive contributions to the thermal partition function. The radius
$R_+$ sets the temperature of the system: $2\pi
T=1/R_+$. The form of the thermal coupling in equation (\ref{untthermal})
is similar to the one that appears in the familiar flat type II
superstring theories at finite temperature. The difference here is
that the temperature is {\em fixed}. Since we have succeeded to
factorize out the thermal lattice, we can now treat all other radii
parameterizing the $\Gamma_{9,9}$ lattice as independent moduli. To
obtain the four dimensional interpretation we discussed in section
$2$, we take the radii $R_{y,z}$ to be large keeping all other ones
small.

As we already remarked, there are only two choices consistent with
the cosmological interpretation of the partition function
corresponding to the two values of the radius $R_+$. For the
diagonal choice we have a radius $R_+^2=1$ corresponding to a temperature $2\pi T=1/R_+$
higher than Hagedorn: $2\pi T_H=1/R_H=1/\sqrt{2}$. This model is
tachyonic and so unstable in perturbation theory. For the second
anti-diagonal choice $R_+^2=4$, and the temperature is {\em below}
Hagedorn: $2\pi T=1/2 <2\pi T_H$. 
This is precisely the temperature that
we gathered from general arguments,
equation (\ref{temperature}) for level $|k|=2$. 
That model gives rise to a well
defined, integrable partition function and a finite norm for the wave-function
at one loop. The integral is difficult to
perform analytically but it can be estimated. We will not carry out
this computation here.

The remaining part consists of the twisted sectors of the theory, $(h,g)\ne
0$. Here we shall find new stringy phenomena associated with the
fact that we are orbifolding the Euclidean time circle. Since this
is twisted, the
thermal co-cycle has to be extended consistently.
% Before doing that
%it is useful to study in more detail this sector.
It takes the general form, valid for both the untwisted and the twisted
sectors: \be
S\left[^{(q+h),~(\alpha+\bar\alpha)}_{(p+g),~(\beta+\bar\beta)}
\right] =e^{i\pi
\left(\,(p+g)(\alpha+\bar\alpha)\,+\,(q+h)(\beta+\bar\beta)\,\right)}.
\ee
That is, the relevant lattice is augmented by the quantum numbers
$(g,h)$ that label the twisted sectors.
Again, the thermal co-cycle insures that fermions contribute
positively to the partition function. In the twisted sectors, there is no
momentum charge and we can set $(p,q)=(0,0)$.

In the twisted sectors, each of the $\hat c=2$ superconformal blocks
is equivalent to a system described by a free complex boson and a free
complex fermion twisted by $Z_4$. This equivalence implies
topological identities for each $N=2$ twisted superconformal block
\cite{Narain:1986qm}\cite{Kiritsis:1997ca}:
%\cite{Kiritsis-kounnas-Infrared},
%\cite{Kiritsis-kounnas-Spontaneous}:
 \be
 {1\over 2|\eta|^4} \sum_{(\gamma_i,\delta_i)}
 \left|\Theta\left[^{\gamma_i}_{\delta_i}\right]\right|~
 \left|\Theta\left[^{\gamma_i-h} _{\delta_i-g}\right]\right|~
 \left|\Theta\left[^{\gamma_i-{h\over 2}}_{\delta_i-{g\over 2}}\right]\right|^2
= {2^2~{\rm sin}^2({\pi \Lambda(h,g)\over 4})}~{\left| \eta
\right|^2 \over
 \left|\Theta\left[^{1+{h\over2}}_{1+{g\over2}}\right]
 \right|^2}
 \ee
 where $\Lambda(h,g)=\Lambda(g,h)$ depend on the $(h,g)$-twisted sector.
 $\Lambda(h,g)=2$ when $(h,g)$ =(0,2), (2,0) and (2,2) while for the
remaining 12 twisted sectors $\Lambda(h,g)=1$.
Although the above orbifold expressions are derived at the fermionic
point, they remain valid for any other point of the untwisted moduli
space. 

Using the above orbifold identity, the ``twisted" part of the
thermal partition function simplifies to:
$$
Z^{thermal}_{twist}=
 ~{1\over
4}\sum_{(\alpha,\beta, \bar\alpha, \bar\beta)} ~{1\over
4}\sum_{(h,g)\ne (0,0)}~{{\rm Im} \tau }~\Gamma_{2,2}
~e^{i\pi(\alpha+\beta+\epsilon\alpha\beta)}
e^{i\pi(\bar\alpha+\bar\beta+\bar\epsilon\bar\alpha\bar\beta)}
e^{i\pi (\alpha + \bar{\alpha})g + i \pi ( \beta + \bar{\beta}) h}
$$
\be \times ~2^8~{\rm sin}^8({\pi\Lambda(h,g)\over 4}) ~{
\Theta\left[^{\alpha+{h\over2}}_{\beta+{g\over2}}\right]^2
\Theta\left[^{\alpha-{h\over2}}_{\beta-{g\over2}}\right]^2
~\bar\Theta\left[^{\bar\alpha
+{h\over2}}_{\bar\beta+{g\over2}}\right]^2
~\bar\Theta\left[^{\bar\alpha
-{h\over2}}_{\bar\beta-{g\over2}}\right]^2 \over \left|
\Theta\left[^{1+{h\over2}}_{1+{g\over2}}\right]
\Theta\left[^{1-{h\over2}}_{1-{g\over2}}\right]\right|^4 }~. \ee
Furthermore, by using the left- (and right-) Jacobi identities \be
{1\over
2}\sum_{(\alpha,\beta)}~e^{i\pi(\alpha+\beta+\epsilon\alpha\beta)}
\Theta\left[^{\alpha+{h\over2}}_{\beta+{g\over2}}\right]^2
\Theta\left[^{\alpha-{h\over2}}_{\beta-{g\over2}}\right]^2=
-|\epsilon|\Theta\left[^{1+{h\over2}}_{1+{g\over2}}\right]^2
\Theta\left[^{1-{h\over2}}_{1-{g\over2}}\right]^2, \ee the twisted
part of the thermal partition function simplifies further:
 \be Z^{thermal}_{twist}=~{1\over 4}~\sum_{(h,g)\ne (0,0)}~{{\rm Im} \tau
}~\Gamma_{2,2}~2^8~{\rm sin}^8({\pi\Lambda(h,g)\over
4})~~\phi((g,h),(\epsilon,\bar\epsilon)). \label{twistedresult}\ee The factor $\phi$
depends on the initial choice of the left- and right-chirality
coefficients $\epsilon, \bar\epsilon$ \be
\phi((g,h),(\epsilon,\bar\epsilon)) =\left(|\epsilon|
+(1-|\epsilon|)(1-e^{i\pi gh})\right)\left(|\bar\epsilon|
+(1-|\bar\epsilon|)(1-e^{i\pi gh})\right).\label{twistedThermal}\ee
It should be noted that expression (\ref{twistedresult}) has a field theoretic
interpretation in terms of a momentum lattice only. 
%String
%oscillators are absent in the twisted sectors.

If the initial non-thermal model was not supersymmetric,
$|\epsilon|=|\bar\epsilon|=1$, then the number of massless bosons
and fermions will not be equal, $n_b\ne n_f$. For all other choices
$n_b = n_f$. This situation is reflected in the factor
$\phi((g,h),(\epsilon,\bar\epsilon))$, which distinguishes the four
different possibilities. In the non supersymmetric case there is a
non-vanishing contribution to the partition function even in the
absence of the thermal co-cycle. This is equivalent, in field
theory, to the one-loop {\em zero temperature contribution to the
effective action}. This contribution is zero in supersymmetric
theories. In the later cases, the corrections are coming from the
massive thermal bosons and fermions plus a contribution from massless
bosons. We will display here a number of typical examples.
The first
class is when the $\Gamma_{2,2}$ lattice is unshifted by $(h,g)$ and
so it factorizes out from the sum over $(g,h)$.

\subsubsection*{Unshifted $\Gamma_{2,2}$ lattice}
In the unshifted case, the sum over $(g,h)$ can be performed easily
so that the only remaining dependence is that of
$\epsilon,\bar\epsilon$. We obtain
\cite{Dixon:1990pc}\cite{Kiritsis-kounnas-Infrared}, \be
S^{thermal}_{twist}~=~\int _F {d\tau d\bar\tau \over ({\rm Im}
\tau)^2}Z^{thermal}_{twist}=-C[\epsilon,\bar\epsilon]~\log
\left[~|\eta (T)|^4 ~|\eta (U)|^4 ~{\rm Im} T~ {\rm Im} U
\mu^2~\right] \label{Sthermal}\ee where $\mu^2$ is an infrared
cut-off and $T,~U$ parameterize the K\"ahler and complex structure
moduli of the target space torus $T^2$. The coefficient
$C[\epsilon,\bar\epsilon]$ depends on the initial chiralities of the
spinors: \be C[1,1]=240,~~~~C[1,0]=C[0,1]=32,~~~~C[0,0]=64 \,.\ee
Actually, $C[\epsilon,\bar\epsilon]$ is nothing but the number of
the massless bosonic degrees of freedom of the theory.
Equation (\ref{Sthermal}) is invariant under the full target space
T-duality group acting on the $T$ and $U$ moduli separately.

For large volume, ${\rm Im} T\gg 1$, the leading behavior is linear
in ${\rm Im} T\sim R_yR_z$. Assuming $iU\sim R_y/R_z$ fixed and
$\mu^2\sim \gamma /R_yR_z$, we have \be
S^{thermal}_{twist}~=~C[\epsilon, \bar \epsilon]~\left({\pi \over
3}~R_y R_z -\log \left[\,\gamma \,|\eta (U)|^4~ {\rm Im} U
~\right]\right). \label{unshifted} \ee Note that in the large volume
limit, the twisted sector contribution to one-loop amplitude depends both on the K\"ahler and complex structure
moduli.

\subsubsection*{Shifted lattice}
Another illustrating example is when the $Z_4$ action shifts the
$\Gamma_{2,2}$ lattice simultaneously with the twist we described
before. In this case the lattice is replaced by a shifted
$\Gamma_{2,2}$ lattice.
The
$(Lh/2,Lg/2)$ shifted lattice, $L=1,2,3$, is given by
\footnote{For brevity we
have given the expression for the shifted $\Gamma_{1,1}$ lattice but
the generalization to the $\Gamma_{2,2}$ lattice is
straightforward.}:
\be
 \Gamma_{1,1}\left[^{{Lh\over 2}}_{{Lg\over 2}}\right](R)= \sum_{(m,n)}~{R \over
\sqrt{{\rm Im}
 \tau}}
 ~e^{-\pi R^2  {{ |(4m+{Lg})+(4n+ {Lh})\tau}|^2 \over {\rm Im}\tau }}.
\ee
Here, we will examine
in more detail the  $L=1$ case which
corresponds to a 1/4-shifted lattice.

When
$(\epsilon,\bar\epsilon)=(1,1)$, the contribution of the twisted
sector to the partition function becomes
 \cite{Kiritsis:1997ca}\cite{Kiritsis-kounnas-Spontaneous}:
\be S^{thermal}_{twist}~=~240~\int _F {d\tau d\bar\tau \over ({\rm Im}
\tau)^2}~{{\rm Im} \tau }
 \left(~\Gamma_{2,2}[T,U]-{1\over4}\Gamma_{2,2}[4T,4U]~\right).
\ee To obtain the above expression we have used the identities \be
~\sum_{(h,g)}\Gamma_{2,2}\left[^{h / 2}_{g / 2}\right] =
\Gamma_{2,2}(T,U),~~~~~~ \Gamma_{2,2}[^{~0~}_{~0~}]
={1\over4}\Gamma_{2,2}[4T,4U]\ee and we subtracted the contribution
of the untwisted sector, $(g,h)=(0,0)$. Integrating over $\tau$ we
obtain \cite{Kiritsis:1997ca} \cite{Kiritsis-kounnas-Spontaneous}
\be S^{thermal}_{twist}~=~- 60~\log \left[{|\eta (T)|^{16} ~|\eta
(U)|^{16} \over |\eta(4T)|^4 ~|\eta (4U)|^4 }~{{\rm Im} T^3~ {\rm
Im} U^3 \mu^6\over 16}\right]. \ee There is no volume factor in the
large ${\rm Im}T$ limit (and this is generic in the case of freely
acting orbifolds
\cite{Kiritsis:1997ca}\cite{Kiritsis-kounnas-Spontaneous}). So for
large ${\rm Im} T$, (and setting ${\rm Im}T \mu^2\sim \gamma$), we
obtain
 \be
S^{thermal}_{twist}~=- 60~\log \left[\,\gamma {~|\eta (U)|^{16}
\over |\eta (4U)|^4 }~~ {{\rm Im} U^3 \over 16}\right]. \label{shifted}
 \ee
In the large volume limit, $S^{thermal}_{twist}$ 
only depends on the complex structure modulus of the torus.

\subsubsection*{Comments}
The total Euclidean one-loop amplitude in both the shifted and
unshifted lattice cases is given by: \be S^{thermal}~=~S^{thermal}_{unt}~+~
S^{thermal}_{twist}\ee The $S^{thermal}_{unt}$ is nothing but
one-quarter of the thermal partition function of type II superstring theory on
$S^1\times T^9$ with all nine spatial radii arbitrary, while
that of Euclidean time fixed by the temperature: $2\pi T=1/R_+=1/2$. By
exponentiating $S^{thermal}$, with the twisted sector contributions in our examples given by
equations (\ref{unshifted}) and (\ref{shifted}), we obtain the norms
of the corresponding cosmological wave-functions as functions of all moduli.

The difference between the shifted and the ordinary model discussed
previously can be understood as follows. It is known that the freely
acting orbifolds are related to gravitational and gauge field backgrounds
with fluxes \cite{Derendinger:2004jn}. This indicates the different
interpretations of the two cosmological models with shifted and
unshifted $\Gamma_{2,2}$. In the shifted model there are
non-vanishing magnetic fluxes \cite{Derendinger:2004jn} while in the
unshifted case such fluxes are absent.

Let us stress here that the thermal $Z_4$-orbifold described involves a
twisting that leaves two moduli parameterizing a $\Gamma_{2,2}$
lattice, which in the large moduli limit gives us the four
dimensional cosmological model discussed in section 2. Many other
orbifold-like models can be constructed, which may factorize bigger
lattices, admitting a higher dimensional interpretation. In all
those cases, the partition function is computable as a function of
the moduli. However its analytic form in terms of the moduli in
various limits depends crucially on whether the orbifold is freely,
or even partially freely acting (or in other words, on the different
structure of magnetic fluxes). Models based on asymmetric orbifolds
can also be constructed giving rise to a rich family of
calculable models. A more detailed analysis would be interesting so
as to understand the classification of the low level cosmologies, as
well as the characteristic dependence of the norm of the
wave-function on the various features of the large class of models.
To illustrate the above points, we offer one further simple example
based on a $Z_2$ instead of the $Z_4$ orbifold.

\subsubsection*{$Z_2$ orbifold models}

In the $Z_2$ orbifold models ($M=2$ in equation (\ref{orbifold})),
the factorization of the cosmological
CFT factor is not so explicit as it was in the $Z_4$ examples. However
the cosmological interpretation remains the same. 

For the untwisted sector, the
genus-1 contribution
$S^{thermal}_{unt}$ is now one half of the thermal partition
function of type II theory on $S^1\times T^9$.
We proceed to analyze the twisted sector contribution to the genus-1
amplitude.
Following similar
steps as in the $Z_4$ orbifold case, and now setting $(2H,
2G)=(h,g)$ to be integers defined modulo $2$, we obtain for the case
$(\epsilon,\bar\epsilon)=(0,0)$
 \be S^{thermal}_{twist}=\int _F {d\tau d\bar\tau \over
({\rm Im} \tau)^2}~{1\over 2}~\sum_{(h,g)\ne (0,0)}~{2^8~{\rm Im} \tau
}~\Gamma_{2,2}~.\ee To arrive at the result notice that since the
characters $(h,g)$ are defined to be integers modulo $2$, the factor
of $1/4$ in the first line of equation (\ref{partition}) now becomes
$1/2$; no other modifications in this formula are needed. All the other
steps carry through as before. Here also we may classify the models
in two classes; in the first class the lattice is taken to be
unshifted, while in the second class the $\Gamma_{2,2}$ is
half-shifted. We will use the definition:
\be
 \Gamma_{1,1}\left[^{{Lh}}_{{Lg}}\right](R)= \sum_{(m,n)}~{R \over
\sqrt{{\rm Im}
 \tau}}
 ~e^{-\pi R^2  {{ |(2m+{Lg})+(2n+ {Lh})\tau}|^2 \over {\rm Im}\tau }}.
\ee

For the {\it unshifted} $\Gamma_{2,2}$ $Z_2$-model, and when
$(\epsilon,\bar\epsilon)=(0,0)$, we obtain: \be
S^{thermal}_{twist}=-384~\log \left[|\eta (T)|^4 ~|\eta (U)|^4 ~{\rm
Im} T~ {\rm Im} U \mu^2\right]. \ee As in the $Z_4$ models, for
large volume, ${\rm Im} T\gg 1$, the leading behavior is linear in
${\rm Im} T\sim R_yR_z$. Assuming $iU\sim R_y/R_z$ fixed and
$\mu^2\sim \gamma /R_yR_z$ we have: \be S^{thermal}_{twist}
=384~\left({\pi \over 3}~R_y R_z -\log \left[~\gamma~ |\eta (U)|^4~
{\rm Im}U ~\right]\right). %\label{unshiftedZ2} 
\ee

The $Z_2$-model with {\it half-shifted} lattice,
$\Gamma_{2,2}[^{h}_{g}],~(g,h=0,1)$, yields
\cite{Kiritsis:1997ca}\cite{Kiritsis-kounnas-Spontaneous}
\be S^{thermal}_{twist}=384~\int
_F {d\tau d\bar\tau \over ({\rm Im} \tau)^2}~{{\rm Im} \tau }
 \left(~\Gamma_{2,2}[\,T,U]-{1\over2}\Gamma_{2,2}[\,2T,2U]~\right).
\label{Z2TwistShifted}\ee
Here also the contribution coming from the
untwisted sector $(h,g)=(0,0)$, is subtracted. To obtain the above
expression we have used the identities \be
~\sum_{(h,g)}\Gamma_{2,2}[^{\,h\,}_{\,g\,}] =
\Gamma_{2,2}[\,T,U],~~~~\Gamma_{2,2}[^{\,0\,}_{\,0\,}]
={1\over2}\Gamma_{2,2}[\,2T,2U]. \ee
Integrating over $\tau$ we
obtain:
 \be S^{thermal}_{twist}=- 192~\log \left[{|\eta (T)|^{8}
~|\eta (U)|^{8} \over |\eta(2T)|^4 ~|\eta (2U)|^4 }~{{\rm Im} T~ {\rm Im} U
\mu^2\over 4}\right]\label{unshiftedZ2}. \ee There is no volume
factor in the large ${\rm Im}T$ limit.
Thus, for large ${\rm Im} T$, (and setting ${\rm Im}T \mu^2\sim
\gamma$), we obtain for the half-shifted lattice
contribution to $S^{thermal}_{twist}$:
 \be
S^{thermal}_{twist}=- 192~\log \left[{~\gamma~|\eta (U)|^{8} \over |\eta
(2U)|^4 }~~ {Im U \over 4}\right]. \label{shiftedZ2}
 \ee

As we see, we can obtain explicit expressions for the genus-1
approximation to the norm of the wave-function for these particular
$Z_2$-models. The twisted sector
contribution is given explicitly by equations
(\ref{unshiftedZ2}) and (\ref{shiftedZ2}).

In the above family of models, we have always considered a two-dimensional
cosmology at small level. That choice is mainly due
to two obstructions that are difficult (but not necessarily impossible)
to circumvent. One is associated to the
difficulty of continuing from Lorentzian to Euclidean signature in
the presence of electric fluxes. The other is that it is difficult
to construct compact models with positive central charge
deficit (negatively curved Euclidean backgrounds) in string theory, or
alternatively,
a compact version of linear dilaton type models.
We make some further comments on
this in
the next section.

\subsection{Liouville type models}
%For non-compact models, the situation is less clearcut, since we
%cannot think of the cosmology as being truly two-dimensional.
%Nevertheless, we can define the above wave-function and its norm,
%and try to interpret it. Again, there are many models in this class,
%and we concentrate on the particular class of models that is:

Consider string theory cosmological backgrounds based on worldsheet
CFTs of the form  (see e.g.
\cite{kounnas-lust}\cite{Nappi:1992kv}\cite{cornalba}\cite{Elitzur:2002rt}\cite{Berkooz:2002je}\cite{Strominger:2003fn}\cite{Hikida:2004mp}\cite{Toumbas:2004fe}\cite{Nakayama:2006gt}):
\begin{eqnarray}
{SL(2,R)_{-|k|} \over U(1)} \times {SL(2,R)_{|k|+4} \over U(1)} \times K.
\end{eqnarray}
A nice feature of such models is that the combined central charge of
the two $SL(2,R)/U(1)$ factors is independent of $|k|$, and so this
can be taken to be an independent, varying parameter.
% In particular,
%it appears that $|k|$ can be taken to be large rendering the
%sigma-model $\alpha'$ expansion valid.
Although the partition function of the euclidean cigar background is known
\cite{Hanany:2002ev}, we need to deal first with the fact that
for such a
background the cosmological wave-function is non-normalizable due to
the infinite volume of the cigar factor $SL(2,R)_{|k|}/ U(1)$. To
produce a normalizable wave-function we must face the problem of
consistently compactifying this factor, as alluded at the end of the
previous section. Moreover, it would be interesting to obtain a
compactification scheme which leaves $|k|$ a free parameter.
Compactifying the cigar would amount to discretizing its continuous
modes keeping at the same time the unitarity and
the modular invariance of
the torus partition function intact.
%, and more importantly the
%unitarity of the theory.
% Unfortunately it is not known yet how to do
%so.

An interesting aspect of these models is that now the sphere
 contribution to the string partition function is finite since the
 volume of the conformal Killing group
 cancels against the volume of the $SL(2,R)_{|k|}/U(1)$ conformal
 field
theory factor. Nevertheless, we can see that the
 torus contribution dominates (at any finite string coupling)
 %again
due to the volume divergence. If a consistent way of cutting off the volume
of the cigar is found, we could interpret
%It could
 %be interesting to analyze whether a sensible limit exists, in which we take
 %the volume (in string units) to infinity, while scaling the string coupling
 %constant to zero, rendering the one-loop contribution finite (and
the torus contribution as a finite thermal correction to the
tree-level contribution, realizing a stringy version of the
computation in \cite{Brustein:2005yn}\cite{Sarangi:2006eb}\cite{kounnas-partouche}.

A further suggestion for developing our formalism in linear dilaton
spaces, is to view the wave-function of the universe as also
depending on a boundary condition in the space-like linear dilaton
direction of the cigar, in order to obtain a linear dilaton
holographic interpretation \cite{Aharony:1998ub}. Part of the
interpretation of the wave-function of the universe would then be as
in the Hartle-Hawking picture, and part would be holographic.
Finally in these models supersymmetry can be restored asymptotically
in the large $|k|$ limit, making contact with linear dilaton models
in null directions (see e.g. \cite{Craps:2005wd} for recent
progress).

\section{Discussion}
We have outlined a framework generalizing the Hartle-Hawking no
boundary proposal of the wave-function of the universe to string theory
cosmological backgrounds.
The class of example cosmologies considered here are described by worldsheet
conformal field theories of the general form $SL(2,R)_{-|k|}/U(1)~\times K$,
where $K$ is an internal, compact CFT.
%These are the only examples we know
%to which the Hartle-Hawking prescription can be generalized.
In order to define the analogue of the Hartle-Hawking wave-function, we
had to surmount the technical hurdle of realizing that such
cosmologies (like the corresponding Euclidean parafermion theories) have
an almost geometrical description in terms of a compact non-singular T-fold.
We then defined
the wave-function of the universe
via a Euclidean string field theory path integral (generalizing the
no-boundary proposal).
For specific examples we computed the norm of the wave-function to leading
order in string perturbation theory,
as a function of moduli parameters. There are many interesting similar examples to
which we can generalize our analysis.

 In a probabilistic interpretation, with a normalizable wave-function at
hand, one can attempt to compute vacuum expectation values for particular physical
quantities in various cosmological models, and analyze
their properties in various
regions of the moduli space.  Our purpose in this paper was to
provide the framework for such a discussion, which promises to be
interesting. In particular, it is an open problem to identify
preferred regions in the moduli space in the large class of models to
which our analysis applies.

More concretely, we believe our construction points out the good use
that can be made of T-folds, and generalized geometry, in stringy
cosmologies (allowing to evade various no-go theorems in pure
geometry). Moreover, we have been able to define a sensible
calculation in a de Sitter-like compactification of string theory,
after analytically continuing to the Euclidean theory. These
calculations are generically hard to come by in de Sitter gravity
%\cite{Antoniadis:1985pj}\cite{Banks:2005bm}
after quantization, so any well-defined cosmological quantity, like
the norm of the wave-function of stringy universes, merits scrutiny.
Finally, we calculated a quantity akin to an entanglement entropy in
de Sitter space, and showed that it only gets contributions starting at one
loop, and we gave its microscopic origin.

\vspace{1.3cm}

\section*{Acknowledgements}

We thank Constantin Bachas, James Bedford, Ben Craps, John
Iliopoulos, Dieter Luest, Herv\'e Partouche, Anastasios Petkou,
Giuseppe Policastro and Marios Petropoulos for useful discussions.
N. T. thanks the Ecole Normale Sup\'erieure and C. K. and J. T.
thank the University of Cyprus for hospitality. This work was
supported in part by the EU under the contracts MRTN-CT-2004-005104,
MRTN-CT-2004-512194 and ANR (CNRS-USAR) contract No 05-BLAN-0079-01
(01/12/05).

\end{document}